\documentclass[a4paper, 11pt]{article}
\usepackage[pdftex, a4paper]{geometry}
\usepackage{graphicx}
\usepackage{subcaption}

\usepackage{amsmath}
\usepackage{amsfonts}
\usepackage{dsfont}
\usepackage{mathrsfs}
\usepackage{amssymb}
\usepackage{bbm}
\usepackage{epsfig}
\usepackage[english]{babel}
\usepackage[all]{xy}
\usepackage{color}

\newcommand*\mcapinn[2]{\vcenter{\hbox{$\mathsurround=0pt
  \ifx\displaystyle#1\textstyle\else#1\fi\bigcap$}}}

\newcommand*\mcupinn[2]{\vcenter{\hbox{$\mathsurround=0pt
  \ifx\displaystyle#1\textstyle\else#1\fi\bigcup$}}}

\DeclareFontFamily{OT1}{pzc}{}
\DeclareFontShape{OT1}{pzc}{m}{it}{<-> s * [1.200] pzcmi7t}{}
\DeclareMathAlphabet{\mathpzc}{OT1}{pzc}{m}{it}

\newtheorem{theorem}{Theorem}
\newtheorem{definition}{Definition}

\newtheorem{lemma}{Lemma}

\newtheorem{proposition}{Proposition}
\newtheorem{assumption}{Assumption}

\title{\bf  Dynamics of Opinions with Social Biases}

\author{Zihan Chen, Jiahu Qin\thanks{J. Qin is with the Department of Automation,
University of Science and Technology of China, Hefei
230027, China (E-mail: jhqin@ustc.edu.cn).}, Bo Li\thanks{B. Li is with the Key Lab of Mathematics Mechanization,  Chinese Academy of Sciences, Beijing 100190,
China (Email: libo@amss.ac.cn).}, Hongsheng Qi\thanks{Z. Chen and H. Qi are with the Key Laboratory of Systems and Control, Academy of Mathematics and Systems Science, Chinese Academy of Sciences, Beijing 100190, China (Email: chenzihan, qihongsh@amss.ac.cn).},   Peter Buchhorn, and Guodong Shi\thanks{P. Buchhorn and G. Shi are with the Research School of Engineering, The Australian National University, ACT 0200, Canberra, Australia    (Email:  peter.buchhorn, guodong.shi@anu.edu.au).}
}

\date{}

\begin{document}

\maketitle

\begin{abstract}
This paper aims to provide  a systemic analysis to   social opinion dynamics subject to individual biases. As a generalization of the classical DeGroot social interactions, defined by  linearly coupled dynamics of peer opinions that evolve over time, biases add to state-dependent edge weights and therefore lead to highly nonlinear network dynamics. Previous studies have dealt with convergence and stability analysis of such systems for a few specific initial node opinions and network structures, and here we focus on how individual biases affect social equilibria and their stabilities. First of all, we prove that when the initial network opinions are polarized towards one side of the state space,  node biases will drive the opinion evolution to the corresponding interval boundaries. Such  polarization attraction  effect continues  to hold under even directed and switching network structures. Next, for a few fundamental network structures, some important  interior network equilibria are presented explicitly for a wide range of system parameters, which are  shown to be locally unstable in general.  Particularly, the interval centroid is proven to be   unstable regardless   of the bias level and the network topologies.
\end{abstract}


\section{Introduction}

\subsection{Background}
Understanding how  opinions of  the members in our society  evolve during their interactions that take place online or in daily lives is becoming increasingly important in many aspects ranging from political decisions to marketing strategies \cite{easley2010networks,friedkin2016network,jackson2010social,scott2017social}. In various cases, social opinions can be represented by real numbers, and by individuals averaging those numbers with neighbors the  classical DeGroot's model was established \cite{degroot1974reaching}. When the social network structure
admits sufficient connectivity, it was shown that DeGroot type of social interactions often leads to convergence to a common opinion, namely agreement or consensus,  across the entire society \cite{tsitsiklis1986distributed,jadbabaie2003coordination,moreau-2005}. The significance of social agreement can be made clear through the notion of naive learning in the sense that a social agreement, even not at the perfect average, implies asymptotic learning of a hidden variable with sufficiently flat interconnections, when nodes' opinions are independently sampled in the first place \cite{naivelearning}.

In practical social networks, however, DeGroot social interactions are arguably rare since it is difficult to observe  social agreement \cite{bias,lawrence2010self,friedkin2016network}.  As a result, a number of generalized models were proposed to capture different psychological effects behind social interactions. Peers might put weight on their initial opinions throughout the entire social interactions as memory effects \cite{friedkin1990social}; nodes might only interact with peers that hold opinions within a given range compared to their own opinions \cite{hegselmann2002opinion,JSAC-Karuse}; a portion of nodes may be stubborn who never revise their initial beliefs \cite{acemouglu2013opinion,yildiz2013binary}; nodes may tend to be repulsive towards enemies by carrying out negative interactions  \cite{altafini1,altafini2013,shi-jsac2013,shi-or2016}. It turned out, beyond asymptotic stability, social dynamics can exhibit complex behaviors such as clustering and oscillation \cite{blondel2009krause,altafini2013,shi-or2016}, being consistent with studies from social and political science \cite{mccarty2016polarized,Plos2010}. In fact, nonlinear bifurcations can arise from collective dynamics of interconnected agents  as a way of gaining survival advantage \cite{naomi2014}.

\subsection{The Model}
Consider a social network with $n$ individuals (nodes) indexed in the set $\mathrm{V}=\{1,\dots,n\}$. The structure of the social network is represented by an undirected graph  $\mathrm{G} = (\mathrm{V},\mathrm{E})$, where each edge $\{i,j\}\in\mathrm{E}$ is an unordered pair of two different nodes in the set $\mathrm{V}$. The graph $\mathrm{G}$ is assumed to be connected without loss of generality. Each $i\in \mathrm{V}$ holds an opinion $\mathbf{x}_{i}(t)\in \mathbb{R}$ at slotted time $t = 0,1,2,...$. Node $i$ interacts with the neighbors in the set $\mathrm{N}_i:=\big\{ \{i,j\}\in\mathrm{E}\big\}$. The influence strength between two neighboring nodes $i$ and $j$ is represented by $w_{ij}>0$ and then $d_{i}:=\sum_{j \in \mathrm{N}_{i} }w_{ij}$ is the total weight of influence applied to node $i$. Note that with connectivity, $\mathrm{N}_i$ is non-empty for any $i$ and thus $d_i>0,i\in\mathrm{V}$. The node $i$'s self-confidence is represented by $w_{ii}>0$. Let $$
\mathbf{s}_{i}(t):=\sum_{j \in \mathrm{N}_i }w_{ij}\mathbf{x}_{j}(t)
$$
be the   external evidence  received by node $i$ at time $t$. Let $b_i$ be a positive number associated with node $i$ as a bias index. The evolution of the  $\mathbf{x}_{i}(t),\ i\in\mathrm{V}$ is described as follows:
\begin{align}\label{sys}
\mathbf{x}_{i}(t+1)=\frac{w_{ii}\mathbf{x}_{i}(t)+\mathbf{x}_{i}^{b_i}(t)\mathbf{s}_{i}(t)}{w_{ii}+\mathbf{x}_{i}^{b_{i}}(t)\mathbf{s}_{i}(t)+
\big(1-\mathbf{x}_{i}(t)\big)^{b_{i}}\big(d_{i}-\mathbf{s}_{i}(t)\big)}.
\end{align}

This model, introduced in \cite{dandekar2013biased}, describes the bias of node $i$ towards the external evidence  $\mathbf{s}_{i}(t)$ compared to its own opinion $\mathbf{x}_i(t)$, as reflected in their respective weights in the update rule. The initial values $\mathbf{x}_i(0),i\in\mathrm{V}$ are assumed to be in the interval $[0,1]$. The level of bias associated with node $i$ is reflected in the value of $b_i$ as a smaller  $b_i$ indicates less biased opinion evolution in the sense that the node dynamics becomes closer to the DeGroot model. It is
easy to verify by induction  that $\mathbf{x}_i(t)\in[0,1]$ for all time instants onwards. However, as an intriguing generalization to the DeGroot model, the high nonlinearity in (\ref{sys}) imposes
fundamental obstacles in establishing further characterizations on the asymptotical behavior of the network dynamics, where only limited results were obtained for very special initial values and network structures    \cite{lawrence2010self,dandekar2013biased}.

\subsection{Contributions}
In this paper, we attempt to provide  a systemic analysis to the social opinion dynamical model (\ref{sys}) with a focus on how individual biases affect social equilibria and their stabilities. First of all, we prove that when the initial network opinions are polarized towards one side of the state interval, such polarization will be
persisted and amplified by node biases during the opinion evolutions in the sense that all node states will converge to the corresponding interval boundaries. Such  polarization attraction  is shown to exist under even directed and switching network structures. Next, we investigate the bias-induced equilibria of the collective nonlinear network dynamics. For fundamental network structures such as complete, star, and cycle graphs, the equilibria are presented explicitly for a wide range of system parameters. The given equilibria are also shown to be locally unstable in general.  Particularly, the interval centroid is shown to be always unstable regardless the choice of bias level and network topologies. These results add to new understandings of the stability analysis in \cite{dandekar2013biased}, going beyond specific initial node opinions despite the high nonlinearity of the network dynamics.

The remainder of the paper is organized as follows. Section \ref{sec:polarization} discusses the polarization attraction effect including the generalizations to directed and switching network stuctures. Section \ref{sec:equilibria} moves on to investigate the new equilibria that arise from the nonlinear network dynamics for both their positions and stabilities. Finally some concluding remarks are given in Section \ref{sec:conclusions}.

\medskip

\noindent{\bf Notation}.
For a vector $X=(x_{1},...,x_{n})^\top \in \mathbb{R}^{n}$, we use $\lVert X \lVert$ to denote its 2-norm, i.e., $\lVert X \lVert=\sqrt{\sum_{i=1}^{n}x_{i}^{2}} $.
For any $x\in \mathbb{R}$, $\lfloor x \rfloor$ represents the largest integer that is no larger than $x$, and $\lceil x \rceil$ represents the smallest integer that is no smaller than $x$.

 \section{Polarization Attraction}\label{sec:polarization}

In this section, we establish the polarization effect of the system (\ref{sys}) when individual opinions are collectively polarized towards one side of the opinion space.

\subsection{Exponential Polarization}

We present the following result.
\begin{theorem}\label{theorem1}
Let $b_{i}>0$ for all $i\in \mathrm{V}$.
\begin{itemize}
 \item[(i)] Suppose $ \mathbf{x}_{i}(0) \in [0,1/2) $ for all $i \in \mathrm{V}$. Then $\lim_{t\to \infty}\mathbf{x}_{i}(t)=0$  for all $ i\in \mathrm{V}$ with
\begin{align*}
\mathbf{x}_{i}(t)\leq \Big( 1-\frac{\alpha}{2}\Big)^{t} \max\limits_{j \in \mathrm{V}}\mathbf{x}_{j}(0),
\end{align*}
where $\alpha= \min\limits_{k \in \mathrm{V}}\frac{d_{k}}{w_{kk}+d_{k}}\Big[\big(1-\max\limits_{j \in \mathrm{V}}\mathbf{x}_{j}(0)\big)^{b_{k}}-\big(\max\limits_{j \in \mathrm{V}}\mathbf{x}_{j}(0)\big)^{b_{k}}\Big]\in (0,1]$.

\item[(ii)] Suppose $\mathbf{x}_{i}(0)\in (1/2,1]$ for all $i\in \mathrm{V}$. Then $\lim_{t\to \infty}\mathbf{x}_{i}(t)=1$  for all $i\in \mathrm{V}$ with
\begin{align*}
\big|\mathbf{x}_{i}(t)-1\big| \leq \Big(1-\frac{\beta}{2}\Big)^{t}\big|\min\limits_{j \in \mathrm{V}}\mathbf{x}_{j}(0)-1\big|,
\end{align*}
where $\beta = \min\limits_{k \in \mathrm{V}}\frac{d_{k}}{w_{kk}+d_{k}}\Big[\big(\min\limits_{j \in \mathrm{V}}\mathbf{x}_{j}(0)\big)^{b_{k}}-\big(1-\min\limits_{j \in \mathrm{V}}\mathbf{x}_{j}(0)\big)^{b_{k}}\Big]\in (0,1]$.
\end{itemize}
\end{theorem}
\medskip

\noindent{\it Proof.}
We consider result (i) at first and divide its proof into two steps.

\noindent Step $1$. Let $\mathbf{y}(t)=\max \limits_{i \in \mathrm{V}}\big\{\mathbf{x}_{i}(t)\big\}$. In this step, we prove that  $\mathbf{y}(t)$ is decreasing. We define
\begin{align*}
\begin{split}
 f_{1}^{i}(t)
 = {} &
w_{ii}+\mathbf{x}_{i}^{b_{i}}(t)\mathbf{s}_{i}(t)+\big(1-\mathbf{x}_{i}(t)\big)^{b_{i}}\big(d_{i}-\mathbf{s}_{i}(t)\big) ,
\\
f_{2}^{i}(t)
= {} &
w_{ii}\big(\mathbf{x}_{i}(t)-\mathbf{y}(t)\big) ,
\\
f_{3}^{i}(t)
={} &
\mathbf{x}_{i}^{b_{i}}(t)\mathbf{s}_{i}(t)\big(1-\mathbf{y}(t)\big)- \big(1-\mathbf{x}_{i}(t)\big)^{b_{i}}\big(d_{i}-\mathbf{s}_{i}(t)\big)\mathbf{y}(t) ,
\\
f_{4}^{i}(t)
={} &
\mathbf{y}(t)\big(1-\mathbf{y}(t)\big)\Big[\mathbf{x}_{i}^{b_{i}}(t)- \big(1-\mathbf{x}_{i}(t)\big)^{b_{i}}\Big] .
\end{split}
\end{align*}
For the $f_{j}^{i}(t),j=1,2,3,4$, the following facts can be established.
\begin{itemize}
\item[(a)] From
 \begin{align*}
 \mathbf{s}_{i}(t)&=\sum_{j \in \mathrm{N}_{i} }w_{ij}\mathbf{x}_{j}(t)\leq \sum_{j \in \mathrm{N}_{i} }w_{ij}\mathbf{y}(t) =d_{i}\mathbf{y}(t)
 \end{align*}
 and $\mathbf{x}_{i}(t)\leq \mathbf{y}(t)<1/2$, there holds that $f_{1}^{i}(t)>0$.
\item[(b)] The definition of $\mathbf{y}(t)$ implies that $f_{2}^{i}(t)\leq 0$.
\item[(c)] If $0 \leq \mathbf{s}_{i}(t)\leq d_{i}\mathbf{y}(t)$, there holds that $f_{3}^{i}(t)\leq d_{i}f_{4}^{i}(t)$.
\item[(d)] If $b_{i}>0$ and $\mathbf{y}(t)<1/2$ hold, we obtain that $\mathbf{x}_{i}(t) \leq \mathbf{y}(t)< 1/2$ and $f_{4}^{i}(t)\leq 0$.
\end{itemize}

From system (\ref{sys}), for all $i\in \mathrm{V}$ and $\mathbf{y}(t) < 1/2$, we obtain
\begin{align}\label{x(t+1)<y(t)}
 &
\mathbf{x}_{i}(t+1) - \mathbf{y}(t)  \nonumber \\
=&
\frac{w_{ii}\mathbf{x}_{i}(t)+\mathbf{x}_{i}^{b_{i}}(t)\mathbf{s}_{i}(t)-w_{ii}\mathbf{y}(t)-\mathbf{x}_{i}^{b_{i}}(t)\mathbf{s}_{i}(t)\mathbf{y}(t)-\big(1-\mathbf{x}_{i}(t)\big)^{b_{i}}\big(d_{i}-\mathbf{s}_{i}(t)\big)\mathbf{y}(t)}
{w_{ii}+\mathbf{x}_{i}^{b_{i}}(t)\mathbf{s}_{i}(t)+\big(1-\mathbf{x}_{i}(t)\big)^{b_{i}}\big(d_{i}-\mathbf{s}_{i}(t)\big)}   \nonumber\\
= &
\frac{f_{2}^{i}(t)+f_{3}^{i}(t)}
{f_{1}^{i}(t)}   \nonumber\\
\leq  &
\frac{f_{2}^{i}(t)+d_{i}f_{4}^{i}(t)}
{f_{1}^{i}(t)}  \nonumber\\
\leq  &
0,
\end{align}
where the first inequation holds with (a), (c) and the second one holds with (b), (d).
Therefore we have proved that if $\mathbf{y}(t) < 1/2$, there hold
\begin{align*}
\mathbf{x}_{i}(t+1) \leq \mathbf{y}(t)
\end{align*}
for all $i \in \mathrm{V}$ and
\begin{align*}
\mathbf{y}(t+1)=\max \limits_{i \in \mathrm{V}}\big\{\mathbf{x}_{i}(t+1)\big\} \leq \mathbf{y}(t) < 1/2.
\end{align*}
Hence, when $\mathbf{y}(0) < 1/2$, we conclude that $\mathbf{y}(t)<1/2$ for all $t \ge 0$ and $\big\{\mathbf{y}(t)\big\}$ is decreasing.

\noindent Step $2$. We will prove that $\big\{\mathbf{y}(t)\big\}$ converges to zero and establish a bound of the convergence rate.
From (\ref{x(t+1)<y(t)}), we know
\begin{align*}
\begin{split}
\mathbf{y}(t)-\mathbf{x}_{i}(t+1)
= {} &
-\frac{f_{2}^{i}(t)+f_{3}^{i}(t)}
{f_{1}^{i}(t)}
\\
\ge {} &
-\frac{1}{w_{ii}+d_{i}}\big(f_{2}^{i}(t)+f_{3}^{i}(t)\big)
\\
\ge {} &
-\frac{d_{i}}{w_{ii}+d_{i}}f_{4}^{i}(t),
\end{split}
\end{align*}
where the first inequation holds due to the facts that $\mathbf{x}_{i}^{b_{i}}(t) \leq 1$ and $\big(1-\mathbf{x}_{i}(t)\big)^{b_{i}} \leq 1$, while the second inequation holds in view of the fact that $f_{2}^{i}(t)\leq 0$ for all $i\in \mathrm{V}$ and $t\ge 0$.

Because $\mathbf{x}_{i}(t) \leq \mathbf{y}(t) \leq \mathbf{y}(0) < 1/2$, there hold
\begin{align*}
\big(1-\mathbf{x}_{i}(t)\big)^{b_{i}}-\mathbf{x}_{i}^{b_{i}}(t) \ge \big(1-\mathbf{y}(0)\big)^{b_{i}}-\mathbf{y}^{b_{i}}(0) >0,
\end{align*}
and
\begin{align*}
-f_{4}^{i}(t)=-\mathbf{y}(t)\big(1-\mathbf{y}(t)\big)\Big[\mathbf{x}_{i}^{b_{i}}(t)- \big(1-\mathbf{x}_{i}(t)\big)^{b_{i}}\Big] \geq \Big[\big(1-\mathbf{y}(0)\big)^{b_{i}}-\mathbf{y}^{b_{i}}(0)\Big]\mathbf{y}(t)\big(1-\mathbf{y}(t)\big).
\end{align*}
This therefore gives us
\begin{align}\label{forall}
\mathbf{x}_{i}(t+1)\leq \mathbf{y}(t)- \frac{d_{i}}{w_{ii}+d_{i}}\Big[\big(1-\mathbf{y}(0)\big)^{b_{i}}-\mathbf{y}^{b_{i}}(0)\Big]\mathbf{y}(t)\big(1-\mathbf{y}(t)\big).
\end{align}
Introduce $\alpha= \min\limits_{k \in \mathrm{V}}\frac{d_{k}}{w_{kk}+d_{k}}\Big[\big(1-\mathbf{y}(0)\big)^{b_{k}}-\mathbf{y}^{b_{k}}(0)\Big]$. Obviously $0<\alpha \leq 1$. Because  $\mathbf{y}(t+1)=\max \limits_{i \in \mathrm{V}}\big\{\mathbf{x}_{i}(t+1)\big\}$ and (\ref{forall}) holds for all $i\in \mathrm{V}$, we obtain
\begin{align*}
\mathbf{y}(t+1)
\leq \mathbf{y}(t)-\alpha \mathbf{y}(t)\big(1-\mathbf{y}(t)\big)
= (1-\alpha)\mathbf{y}(t)+\alpha \mathbf{y}^{2}(t)
\leq(1-\alpha)\mathbf{y}(t)+\frac{\alpha}{2} \mathbf{y}(t)
= \Big(1-\frac{\alpha}{2}\Big)\mathbf{y}(t).
\end{align*}
Therefore, for all $i \in \mathrm{V}$,
\begin{align*}
\mathbf{x}_{i}(t)\leq \mathbf{y}(t) \leq \bigg(1-\frac{\alpha}{2}\bigg)^{t}\mathbf{y}(0)=\bigg(1-\frac{\alpha}{2}\bigg)^{t} \max\limits_{j \in \mathrm{V}}\mathbf{x}_{j}(0).
\end{align*}
This proves (i). holds.
The statement (ii) follows from a similar argument, whose details are omitted.
Now we have completed the proof.  \hfill$\Box$

Note that Theorem \ref{theorem1} demonstrates the fundamental difference between the DeGroot type of social interactions  and the nonlinear  opinion dynamics (\ref{sys}). Particularly, DeGroot model defines contraction mappings in the opinion space \cite{tsithesis,blondel2005,caoming2008}, where the metric $$
\max_{i\in\mathrm{V}} \mathbf{x}_i(t)-\min_{i\in\mathrm{V}} \mathbf{x}_i(t)
$$
is monotonically decreasing for any network structure. With a fixed interaction structure,  convergence of DeGroot model can be explained by spectrum of the state transition matrix from standard linear systems theory \cite{xiao}, however, the contraction nature of the DeGroot dynamics  is certainly beyond that which holds true even under random node interactions \cite{jackson2010social,shi-tit-2015} or nonlinear edge weights \cite{moreau-2005,bauso2006,lin2007}. The proof of Theorem \ref{theorem1} illustrates that $\max_{i\in\mathrm{V}} \mathbf{x}_i(t)$ is no longer contracting along (\ref{sys}). Instead, when $\max_{i\in\mathrm{V}} \mathbf{x}_i(t)<1/2$, the entire network dynamics will be pushed to the boundary of the opinion space.
\subsection{Directed and Switching Graph}

We now generalize Theorem \ref{theorem1} to networks with directed and switching structures. To this end, let $\mathrm{G}(t) = \big(\mathrm{V},\mathrm{E}(t)\big)$ be a time-varying directed graph where at time $t$, the edge set $\mathrm{E}(t)$ consists of some directed arcs as ordered pairs from the set $\mathrm{V}$. Node $i$'s self-confidence at time $t$ is $w_{ii}(t)$, and the arc $(j,i)\in\mathrm{E}(t)$ holds a weight $w_{ij}(t)$. The neighbor set of node $i$ at time $t$ is in turn defined as $$
\mathrm{N}_i(t):=\big\{j:(j,i)\in\mathrm{E}(t) \big\}.
$$
Let $\mathbf{s}_{i}(t):=\sum_{j \in \mathrm{N}_i(t)}w_{ij}(t)\mathbf{x}_{j}(t)$ and $d_{i}(t):=\sum_{j \in \mathrm{N}_i(t) }w_{ij}(t)$. The network dynamics becomes
\begin{align}\label{sys2}
\mathbf{x}_{i}(t+1)=\frac{w_{ii}(t)\mathbf{x}_{i}(t)+\mathbf{x}_{i}^{b_{i}}(t)\mathbf{s}_{i}(t)}{w_{ii}(t)+\mathbf{x}_{i}^{b_{i}}(t)\mathbf{s}_{i}(t)+\big(1-\mathbf{x}_{i}(t)\big)^{b_{i}}\big(d_{i}(t)-\mathbf{s}_{i}(t)\big)}, \ i\in\mathrm{V}.
\end{align}

We impose the following assumption.
\begin{assumption}The following hold for the system (\ref{sys2}).
\begin{itemize}
\item[(i)] there exist  $\mathsf{w}_{ii}\ge 0,i\in \mathrm{V}$ such that  $w_{ii}(t)\leq \mathsf{w}_{ii}$ for all $t\ge 0$ and all $i\in \mathrm{V}$;
\item[(ii)] there exits $c>0$ such that $d_{i}(t)\ge c$ whenever $d_{i}(t)>0$ for all $i\in\mathrm{V}$;
 \item[(iii)]  there is $T \in \mathbb{N}^{+}$ such that $\sum_{s=t}^{t+T-1}d_{i}(s)>0$ for any $t\ge 0$ and $i\in \mathrm{V}$.
    \end{itemize}
    \end{assumption}
It turns out, the polarization effect continues to exist under this directed and time-varying node interactions, as shown in the following result.

\begin{proposition}\label{switching}
 Suppose Assumption 1 holds. Then the following statements hold true.

\begin{itemize}
\item[(i)] If $ \mathbf{x}_{i}(0) \in [0,1/2) $ for all $i \in \mathrm{V}$, then $\lim_{t\to \infty}\mathbf{x}_{i}(t)=0$ for all  $i\in \mathrm{V}$ with
\begin{align*}
\mathbf{x}_{i}(t)\leq \Big(1-\frac{\alpha_{*}}{2}\Big)^{\lfloor t/T \rfloor}\max\limits_{j \in \mathrm{V}}\mathbf{x}_{j}(0),
\end{align*}
where $\alpha_{*}= \min\limits_{k \in \mathrm{V}}\frac{c}{\mathsf{w}_{kk}+c}\Big[\big(1-\max\limits_{j \in \mathrm{V}}\mathbf{x}_{j}(0)\big)^{b_{k}}-\big(\max\limits_{j \in \mathrm{V}}\mathbf{x}_{j}(0)\big)^{b_{k}}\Big]\in (0,1]$.

\item[(ii)] If $\mathbf{x}_{i}(0)\in (1/2,1]$ for all $i\in \mathrm{V}$, then $\lim_{t\to \infty}\mathbf{x}_{i}(t)=1$ for all $i\in \mathrm{V}$ with
\begin{align*}
\big|\mathbf{x}_{i}(t)-1\big| \leq \Big(1-\frac{\beta_{*}}{2}\Big)^{\lfloor t/T \rfloor}\big|\min\limits_{j \in \mathrm{V}}\mathbf{x}_{j}(0)-1\big|,
\end{align*}
where $\beta_{*} = \min\limits_{k \in \mathrm{V}}\frac{c}{\mathsf{w}_{kk}+c}\Big[\big(\min\limits_{j \in \mathrm{V}}\mathbf{x}_{j}(0)\big)^{b_{k}}-\big(1-\min\limits_{j \in \mathrm{V}}\mathbf{x}_{j}(0)\big)^{b_{k}}\Big]\in (0,1]$.
\end{itemize}
\end{proposition}

\noindent{\it Proof.}
(i). We continue to use the definition $\mathbf{y}(t)=\max \limits_{i \in \mathrm{V}}\big\{\mathbf{x}_{i}(t)\big\}$. Furthermore, we define
\begin{align*}
\tilde{\mathbf{y}}(m)=\max\limits_{mT\leq h\leq (m+1)T-1}\big\{\mathbf{y}(h)\big\}=\max\limits_{\substack{mT\leq h\leq (m+1)T-1,\\ i\in \mathrm{V}}}{\mathbf{x}_{i}(h)},
\end{align*}
where $m \in \mathbb{N}$.
For all $i \in \mathrm{V}$, if $d_{i}(t)=s_{i}(t)=0$, there holds $\mathbf{x}_{i}(t+1)=\mathbf{x}_{i}(t)$.
When  $d_{i}(t)>0$ and $\mathbf{y}(t)<1/2$, $\mathbf{x}_{i}(t+1) \leq \mathbf{y}(t)$ from (\ref{x(t+1)<y(t)}). Therefore, from $\mathbf{y}(0) < 1/2$, we conclude that for all $i \in \mathrm{V}$ and $t\ge 0$, $\mathbf{y}(t)<1/2$ for all $t \ge 0$ and $\mathbf{y}(t)$ is decreasing. Then $\tilde{\mathbf{y}}(m)=\mathbf{y}(mT)$ holds and $\tilde{\mathbf{y}}(m)$ is decreasing.

We will prove that $\big\{\tilde{\mathbf{y}}(m)\big\}$ converge to zero and establish the convergence rate.
Let $\tilde{t}\in \big[mT,\infty\big)$ where $m\in \mathbb{N}$ such that $d_{i}(\tilde{t})>c>0$. We see
\begin{align*}
\begin{split}
\mathbf{x}_{i}(\tilde{t}+1)\leq {} &
  \mathbf{y}(\tilde{t})- \frac{d_{i}(\tilde{t})}{w_{ii}(\tilde{t})+d_{i}(\tilde{t})}\Big[\big(1-\mathbf{y}(0)\big)^{b_{i}}-\mathbf{y}^{b_{i}}(0)\Big]\mathbf{y}(\tilde{t})\big(1-\mathbf{y}(\tilde{t})\big)
  \\
   \leq {} &
 \mathbf{y}(\tilde{t})- \frac{d_{i}(\tilde{t})}{\mathsf{w}_{ii}+d_{i}(\tilde{t})}\Big[\big(1-\mathbf{y}(0)\big)^{b_{i}}-\mathbf{y}^{b_{i}}(0)\Big]\mathbf{y}(\tilde{t})\big(1-\mathbf{y}(\tilde{t})\big).
\end{split}
\end{align*}
Furthermore, when $t\in[mT,\infty)$, in view of $\frac{d_{i}(t)}{\mathsf{w}_{ii}+d_{i}(t)}\Big[\big(1-\mathbf{y}(0)\big)^{b_{i}}-\mathbf{y}^{b_{i}}(0)\Big] \in (0,1]$ and
 $\tilde{\mathbf{y}}(m)\ge \mathbf{y}(t)$,  there holds
\begin{align*}
\begin{split}
{} &\mathbf{y}(t)- \frac{d_{i}(t)}{\mathsf{w}_{ii}+d_{i}(t)}\Big[\big(1-\mathbf{y}(0)\big)^{b_{i}}-\mathbf{y}^{b_{i}}(0)\Big]\mathbf{y}(t)\big(1-\mathbf{y}(t)\big) \\
\leq {} &
\tilde{\mathbf{y}}(m)- \frac{d_{i}(t)}{\mathsf{w}_{ii}+d_{i}(t)}\Big[\big(1-\mathbf{y}(0)\big)^{b_{i}}-\mathbf{y}^{b_{i}}(0)\Big]\tilde{\mathbf{y}}(m)\big(1-\tilde{\mathbf{y}}(m)\big).
\end{split}
\end{align*}
Therefore, we obtain
\begin{align}\label{period}
\mathbf{x}_{i}(\tilde{t}+1)
\leq &
\tilde{\mathbf{y}}(m)- \frac{d_{i}(\tilde{t})}{\mathsf{w}_{ii}+d_{i}(\tilde{t})}\Big[\big(1-\mathbf{y}(0)\big)^{b_{i}}-\mathbf{y}^{b_{i}}(0)\Big]\tilde{\mathbf{y}}(m)\big(1-\tilde{\mathbf{y}}(m)\big)
\nonumber\\
\leq &
\tilde{\mathbf{y}}(m)- \frac{c}{\mathsf{w}_{ii}+c}\Big[\big(1-\mathbf{y}(0)\big)^{b_{i}}-\mathbf{y}^{b_{i}}(0)\Big]\tilde{\mathbf{y}}(m)\big(1-\tilde{\mathbf{y}}(m)\big).
\end{align}
Because $\mathbf{x}_{i}(t+1)=\mathbf{x}_{i}(t)$ when $d_{i}(t)=0$, we conclude  that
\begin{align}\label{switching1}
\tilde{\mathbf{y}}(m+1)=\max\limits_{\substack{(m+1)T\leq h\leq (m+2)T-1,\\ i\in \mathrm{V}}}{\mathbf{x}_{i}(h)}\leq \max\limits_{\substack{mT+1\leq h\leq (m+2)-1T,\\d_{i}(h)>0, i\in \mathrm{V}}}{\mathbf{x}_{i}(h)}
\end{align}
 for all $i\in \mathrm{V}$.

Introduce $\alpha_{*}= \min\limits_{m \in \mathrm{V}}\frac{c}{\mathsf{w}_{mm}+c}\Big[\big(1-\mathbf{y}(0)\big)^{b_{m}}-\mathbf{y}^{b_{m}}(0)\Big]$. Obviously there holds $0<\alpha_{*} \leq 1$.
 Due to (\ref{period}) and (\ref{switching1}), we thus have
\begin{align*}
\begin{split}
\tilde{\mathbf{y}}(m+1)
\leq  {} &
\tilde{\mathbf{y}}(m)-\alpha_{*} \tilde{\mathbf{y}}(m)\big(1-\tilde{\mathbf{y}}(m)\big)
\\
= {} & (1-\alpha_{*})\tilde{\mathbf{y}}(m)+\alpha_{*} \tilde{\mathbf{y}}^{2}(m)
\\
\leq {} &
(1-\alpha_{*})\tilde{\mathbf{y}}(m)+\frac{\alpha_{*}}{2} \tilde{\mathbf{y}}(m)
\\
= {} &
\Big(1-\frac{\alpha_{*}}{2}\Big)\tilde{\mathbf{y}}(m).
\end{split}
\end{align*}
for all $m\in\mathbb{N}$.
Therefore, for all $i \in \mathrm{V}$, $\tilde{\mathbf{y}}(m) \leq (1-\frac{\alpha_{*}}{2})^{m}\tilde{\mathbf{y}}(0)$.
From the definition of  $\tilde{\mathbf{y}}(m)$, we know
\begin{align*}
\mathbf{x}_{i}(t)\leq \Big(1-\frac{\alpha_{*}}{2}\Big)^{\lfloor t/T \rfloor}\max\limits_{j \in \mathrm{V}}\mathbf{x}_{j}(0).
\end{align*}

\noindent(ii). The statement follows from the same analysis as in the proof of (i). We thus have completed the proof. \hfill$\Box$

\medskip

It is worth emphasizing that in Proposition \ref{switching}, the two conditions (i)-(ii) of Assumption 1 are just technical conditions which are consistent with standard DeGroot consensus algorithms \cite{blondel2005,caoming2008}. On the other hand, the condition
(iii)  of Assumption 1 serves as a {\em connectivity} assumption. However, such connectivity is significantly weaker than the usual connectivity assumptions for DeGroot consensus algorithm in the sense that it only requires each node must be affected by some other node during the series of  bounded time intervals.

\section{The Induced Equilibria}\label{sec:equilibria}
In this section, we investigate the bias-induced equilibria of the system (\ref{sys}).
Clearly, the total number of  degrees of freedom is too high to facilitate a meaningful analysis given the bias levels $b_i$ and the node weights $w_{ij}$. To ease the presentation, we impose the following assumption in this section.

\medskip

\begin{assumption}\label{assumption} The following hold for the system (\ref{sys}):
\begin{itemize}
\item [(i)] there is $b>0$ such that $b_{i}=b$ for all $i\in \mathrm{V}$;
  \item[(ii)] $w_{ij}=1$ for all $\{i,j\}\in \mathrm{E}$;
  \item[(iii)] $w_{ii}=w\geq0$ for all $i\in \mathrm{V}$.
  \end{itemize}
\end{assumption}

\medskip

We assume the above assumption throughout the remainder of this section. Let $\mathbb{E}$ be the set of equilibria of (\ref{sys}) and $\mathbb{E}_{\rm bdy}=\{\mathsf{X} \in [0,1]^{n} \backslash (0,1)^{n}:\mathsf{X} \in \mathbb{E}\}$ be  the set of boundary  equilibria. It is clear that every point in $\{0,1\}^{n}$ is a boundary equilibrium. we introduce
  $\mathbb{E}_{\rm int}=\{\mathsf{X}\in (0,1)^{n} :\mathsf{X}\in \mathbb{E}\}$ as the set of interior equilibria, which is certainly of more interest. Furthermore, denote $\mathbf{X}(t)=\big(\mathbf{x}_{1}(t),\mathbf{x}_{2}(t),...,\mathbf{x}_{n}(t)\big)^\top$. Recall the following definition \cite{khalil2000nonlinear}.

 \begin{definition}\label{stable}
The equilibrium $\mathsf{X}=(\mathsf{x}_{1},\mathsf{x}_{2},...,\mathsf{x}_{n-1},\mathsf{x}_{n})^\top$ of system (\ref{sys}) is locally stable if for every $\varepsilon >0$ there exists a $\delta=\delta(\varepsilon)>0$ such that $\lVert \mathbf{X}(t)-\mathsf{X}\lVert < \varepsilon$ for all $t\ge 0$ whenever $\lVert \mathbf{X}(0) -\mathsf{X}\lVert <\delta$. Otherwise, the equilibrium is called to be unstable.
\end{definition}

\subsection{Equilibria Distribution}

For any equilibrium $\mathsf{X}$, there holds that
\begin{align*}
\mathsf{x}_{i}=\frac{w_{ii}\mathsf{x}_{i}+\mathsf{x}_{i}^{b}\big(\sum_{j \in \mathrm{N}_i }w_{ij}\mathsf{x}_{j}\big)}{w_{ii}+\mathsf{x}_{i}^{b}\big(\sum_{j \in \mathrm{N}_i }w_{ij}\mathsf{x}_{j}\big)+
\big(1-\mathsf{x}_{i}\big)^{b}\big(d_{i}-(\sum_{j \in \mathrm{N}_i }w_{ij}\mathsf{x}_{j})\big)}, \ i\in\mathrm{V}
\end{align*}
which is equivalent to
\begin{align}\label{eqn:psys}
   p_i(\mathsf{x}_1,\dots,\mathsf{x}_n):= \mathsf{x}_i^b\Big(\mathsf{x}_i-1\Big)\Big(\sum_{j \in \mathrm{N}_i }w_{ij}\mathsf{x}_{j}\Big)+\mathsf{x}_i\Big(1-\mathsf{x}_i\Big)^{b}\Big(d_{i}-\big(\sum_{j \in \mathrm{N}_i }w_{ij}\mathsf{x}_{j}\big)\Big)=0, \ i\in\mathrm{V}.
\end{align}
Here each $p_i(\mathsf{x}_1,\dots,\mathsf{x}_n)$ is a polynomial function.

We note that there exist methods from computational  algebraic geometry \cite{Cox2015} to find the solutions to (\ref{eqn:psys}). Let $ \mathbb{C}[\mathsf{x}_1,\dots,\mathsf{x}_n]$ denote all the complex polynomials over $n$ variables $\mathsf{x}_1,\dots,\mathsf{x}_n$. For any set of polynomials $\{q_1,\dots, q_s\}$,
\begin{align*}
  \langle q_1,\dots,q_s \rangle_{[\mathsf{x}_1,\dots,\mathsf{x}_n]} = \Big\{\sum_{i=1}^s h_iq_i: h_1,\dots,h_s\in \mathbb{C}[\mathsf{x}_1,\dots,\mathsf{x}_n]\Big\},
\end{align*}
is called the ideal generated by $q_1,\dots,q_s$. We set $
  \mathfrak{I} =\langle p_1,\dots,p_n \rangle_{[\mathsf{x}_1,\dots,\mathsf{x}_n]}$ and define
\[
\mathbb{E}_\mathfrak{I} = \big\{\left(\mathsf{x}_1,\dots,\mathsf{x}_n\right)\in \mathbb{C}^n : \forall g\in\mathfrak{I}, g(\mathsf{x}_1,\dots,\mathsf{x}_n)=0\big\}
\]
as solutions of the ideal $\mathfrak{I}$. It is easy to verify that  $\mathbb{E}_\mathfrak{I}$ is the same as $\mathbb{E}$, and it turns out using  Buchberger's Algorithm, we may be able to  recursively obtain   $\mathbb{E}_\mathfrak{I}$ from solving single variable polynomial equations, during which Sturm's Theorem  helps us find the exact number of solutions in $[0,1]^{n}$. The process of  Buchberger's Algorithm is essentially finding a new generating set of polynomials of $\mathfrak{I}$ which have simpler forms in a similar fashion as Gaussian eliminations.

The $2^n$ equilibria in the set
$
\{0,1\}^n
$
are also quite interesting as they are vertex equilibria in the opinion space. Their stabilities would reflect stubborn and extreme  social formations. We provide the following example.

\medskip

\begin{figure}
\centering
\includegraphics[width=6.2in]{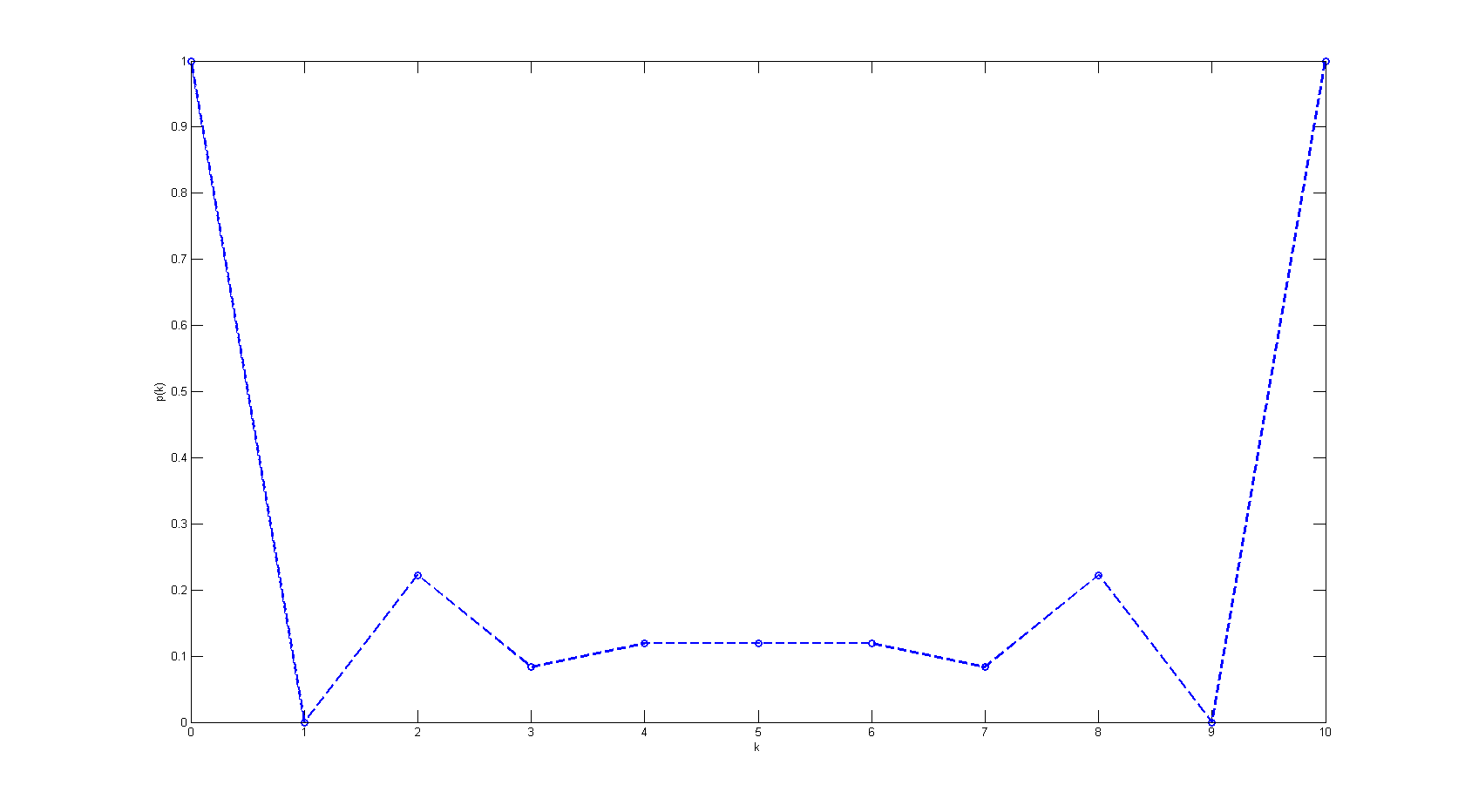}
\caption{The ratio of stable equilibria in the set  $\mathbb{E}_{\rm ver}^k$. }
\label{fig:stability}
\end{figure}

\noindent{\bf Example.} Consider a cycle graph with $10$ nodes subject to Assumption $2$ with $b=3$. The $2^{10}$ vertex equilibria are denoted by $$
\mathbb{E}_{\rm ver}:=\Big\{(\mathsf{x}_1 \ \dots \ \mathsf{x}_{10})^\top: \mathsf{x}_i\in\{0,1\}\Big\}.
$$
The stability of each equilibrium in the set $\mathbb{E}_{\rm ver}$ is tested by randomization method: around each equilibrium a total of $100$ initial values are selected randomly, where each coordinate of these initial values is within $0.015$ compared to the equilibrium; the algorithm is run for $10^4$ steps and if the distance between the resulting outcome and the equilibrium is always within three times of the initial distance for the $100$ initial values, the equilibria is considered as stable. We denote
$$
\mathbb{E}_{\rm ver}^k:=\Big\{(\mathsf{x}_1 \ \dots \ \mathsf{x}_{10})^\top: \mathsf{x}_i\in\{0,1\},\ \sum_{i=1}^{10} \mathsf{x}_i=k\Big\}
$$
for $k=0,1,\dots,10$. The subset of stable equilibria of $\mathbb{E}_{\rm ver}^k$ is denoted by $\overline{\mathbb{E}}_{\rm ver}^k$ We define
$$
p(k)=:{\big|\overline{\mathbb{E}}_{\rm ver}^k\big|}/{\big|\mathbb{E}_{\rm ver}^k\big|}
$$
as the ratio of stable equilibria in the set $\mathbb{E}_{\rm ver}^k$. The plot of $p(k)$ is shown in Figure \ref{fig:stability}.

The numerical result illustrates that in most cases of $k$, both stable and unstable equilibria exist in the set $\mathbb{E}_{\rm ver}^k$. Moreover, $p(k)$ is symmetric with respect to $k=5$, which seems natural in view of the construction of the set $\mathbb{E}_{\rm ver}^k$ and the symmetry of a cycle graph.
 \subsection{Main Results}
 Note that the graph $\mathrm{G}$  is a complete graph if $\{i,j\}\in \mathrm{E}$ for all $i,j\in \mathrm{V}$; a star graph
 if $\mathrm{E}=\big\{\{i,n\}:\ i=1,2,...,n-1. \big\}$; and a cycle graph
 if $\mathrm{E}=\big\{\{1,2\},\{2,3\},...,\{n-1,n\},\{n,1\}\big\}$. We use the convenience that node $nm+k$ represents node $k\in \mathrm{V}$ for all $m\in \mathbb{Z}$.

First of all, it can be easily seen that the opinion space centroid $(1/2,1/2,...,1/2)^\top$ is always an unstable interior equilibrium.

\begin{proposition}\label{1/2 unstable} Let Assumption 2 hold.  Then
$\mathsf{X}=(1/2,1/2,...,1/2)^\top$ is always an unstable equilibrium of system (\ref{sys}).
\end{proposition}

When the underlying network structure is a complete graph, it can be shown that the set of interior equilibria contains the singleton $(1/2,1/2,...,1/2)^\top$ only.

\begin{theorem}\label{complete equilibria}
  Let $\mathrm{G}$ be a complete graph with $n\geq 3$ subject to  Assumption 2. Then    $$
  \mathbb{E}_{\rm int}=\big\{(1/2,1/2,...,1/2)^\top\big\}
  $$ if $b\leq 1$ or $b=2$. Moreover,  the equilibrium  $(1/2,1/2,...,1/2)^\top$ is unstable.\end{theorem}

For star and cycle graphs, a variety of new interior equilibria arises from the nonlinear network dynamics, as presented in the following two results.

\begin{theorem}\label{star equilibria}
Let $\mathrm{G}$ be a star graph  subject to  Assumption 2. Then the following statements hold.
\begin{itemize}
\item[(i)]    $\mathbb{E}_{\rm int}=\big\{(\mathsf{x}_{1},\mathsf{x}_{2},...,\mathsf{x}_{n-1},1/2)^\top: \sum_{i=1}^{n-1}
\mathsf{x}_{i}=(n-1)/2,\mathsf{x}_{i}\in (0,1) \text{ for all } i\in \mathrm{V} \backslash \{n\} \big\}$ if $b=1$;

\item[(ii)]   $\mathbb{E}_{\rm int}=\big\{(\mathsf{x}_{1},\mathsf{x}_{1},...,\mathsf{x}_{1},1-\mathsf{x}_{1})^\top: \mathsf{x}_{1} \in (0,1) \big\}$ if $b=2$;
\item[(iii)]  $\mathbb{E}_{\rm int}=\big\{(1/2,1/2,...,1/2)^\top\big\}$ if $b\neq 1,2$.

\end{itemize}
Moreover, any equilibrium  $\mathsf{X}\in \mathbb{E}_{\rm int}$ is unstable.
\end{theorem}

\begin{theorem}\label{cycle equilibria}
Let $\mathrm{G}$ be a cycle graph  subject to  Assumption 2. Then the following statements hold.
\begin{itemize}
\item[(i)] If $b=1$ and $n\equiv 1,2 \text{ or }3 (mod \ 4)$, $\mathbb{E}_{\rm int}=\big\{(1/2,1/2,...,1/2)^\top\big\}$;
\item[(ii)] If $b=1$ and $n\equiv 0 (mod \ 4)$,
 $\mathbb{E}_{\rm int}=\big\{(a_{1},a_{2},1-a_{1},1-a_{2},a_{1},...,1-a_{1},1-a_{2})^\top:a_{1},a_{2} \in (0,1) \big\}$;
\item[(iii)] If $b=2$ and $n\equiv 1 (mod \ 2)$, $\mathbb{E}_{\rm int}=\big\{(1/2,1/2,...,1/2)^\top\big\}$;
\item[(iv)] If $b=2$ and $n\equiv 0 (mod \ 2)$,  $\mathbb{E}_{\rm int}=\big\{(a,1-a,a,1-a,...,a,1-a)^\top: a \in (0,1) \big\}$.
\end{itemize}
Moreover,  any equilibrium  $\mathsf{X}\in \mathbb{E}_{\rm int}$ is unstable for $b=1$ or $b=2$.
\end{theorem}

It appears to be extremely difficult in generalizing these results to networks with a less common structure. The reasoning comes from that the ideal generated by the polynomials in (\ref{eqn:psys}) depends on the network structure in a highly nontrivial manner. While as we explained above, solving such systems of polynomial equations are equivalent to solving such equations on the generated ideals of the polynomials \cite{Cox2015}. Nonetheless, these results illustrate intriguing equilibria can indeed arise for the system (\ref{sys}). We conjecture that the majority of the interior equilibria should be unstable.


\subsection{Key Lemma}

We define the invariance potential function as follows.
\begin{definition}
Let Assumption \ref{assumption} hold. The invariance potential function of $\mathbf{x}_{i}(t)\in (0,1)$ is defined as
\begin{align*}
\mathbf{s}^{*}\big(\mathbf{x}_{i}(t),b\big)=\frac{\big(1-\mathbf{x}_{i}(t)\big)^{b-1}}{\mathbf{x}_{i}^{b-1}(t)+\big(1-\mathbf{x}_{i}(t)\big)^{b-1}}.
\end{align*}
\end{definition}

We present the following key technical lemma indicating the role of the invariance potential function.
\begin{lemma}\label{lemma 1}
Suppose that $\mathbf{x}_{i}(t)\in (0,1),i\in \mathrm{V}$. Then the following statements hold.
\begin{itemize}
\item[(i)] $\mathbf{x}_{i}(t+1)=\mathbf{x}_{i}(t)$ if and only if $\mathbf{s}_{i}(t)/d_{i}=\mathbf{s}^{*}\big(\mathbf{x}_{i}(t),b\big)$;

\item[(ii)] $\mathbf{x}_{i}(t+1)>\mathbf{x}_{i}(t)$ if  $\mathbf{s}_{i}(t)/d_{i}>\mathbf{s}^{*}\big(\mathbf{x}_{i}(t),b\big)$;

\item[(iii)] $\mathbf{x}_{i}(t+1)<\mathbf{x}_{i}(t)$ if $\mathbf{s}_{i}(t)/d_{i}<\mathbf{s}^{*}\big(\mathbf{x}_{i}(t),b\big)$.
\end{itemize}
\end{lemma}

\noindent{\it Proof.}
(i). Since $\mathbf{x}_{i}(t)\in (0,1)$, there hold $\big(1-\mathbf{x}_{i}(t)\big)\in (0,1), \mathbf{x}_{i}^{b}(t)>0$ and $\big(1-\mathbf{x}_{i}(t)\big)^{b} > 0$. As a result,
\begin{align*}
\begin{split}
{} &
\mathbf{x}_{i}(t)=\mathbf{x}_{i}(t+1)
\\
\Longleftrightarrow {} \ \ &
\mathbf{x}_{i}(t)=\frac{w\mathbf{x}_{i}(t)+\mathbf{x}_{i}^{b}(t)\mathbf{s}_{i}(t)}{w+\mathbf{x}_{i}^{b}(t)\mathbf{s}_{i}(t)+\big(1-\mathbf{x}_{i}(t)\big)^{b}\big(d_{i}-\mathbf{s}_{i}(t)\big)}
\\
\Longleftrightarrow  {}\ \  &
\mathbf{x}_{i}^{b+1}(t)\mathbf{s}_{i}(t)+\mathbf{x}_{i}(t)\big(1-\mathbf{x}_{i}(t)\big)^{b}\big(d_{i}-\mathbf{s}_{i}(t)\big)=\mathbf{x}_{i}^{b}(t)\mathbf{s}_{i}(t)
\\
\Longleftrightarrow  {}\ \  &
\Big[\mathbf{x}_{i}^{b}(t)\big(1-\mathbf{x}_{i}(t)\big)+\mathbf{x}_{i}(t)\big(1-\mathbf{x}_{i}(t)\big)^{b}\Big]\mathbf{s}_{i}(t)=\mathbf{x}_{i}(t)\big(1-\mathbf{x}_{i}(t)\big)^{b}d_{i}
\\
\Longleftrightarrow  {}\ \  &
\frac{\mathbf{s}_{i}(t)}{d_{i}}=\frac{\big(1-\mathbf{x}_{i}(t)\big)^{b-1}}{\mathbf{x}_{i}^{b-1}(t)+\big(1-\mathbf{x}_{i}(t)\big)^{b-1}}=\mathbf{s}^{*}\big(\mathbf{x}_{i}(t),b\big).
\end{split}
\end{align*}
This proves (i).

\noindent(ii). We calculate the partial derivative of $\mathbf{x}_{i}(t+1)$ in system (\ref{sys}), that is,
\begin{align}\label{partial}
  &
\frac{\partial \mathbf{x}_{i}(t+1)}{\partial \mathbf{s}_{i}(t)}
\nonumber\\
= &
\frac{\mathbf{x}_{i}^{b}(t)\Big[w+\mathbf{x}_{i}^{b}(t)\mathbf{s}_{i}(t)+\big(1-\mathbf{x}_{i}(t)\big)^{b}\big(d_{i}-\mathbf{s}_{i}(t)\big)\Big]-\big(w\mathbf{x}_{i}(t)+\mathbf{x}_{i}^{b}(t)\mathbf{s}_{i}(t)\big)\Big[\mathbf{x}_{i}^{b}(t)-\big(1-\mathbf{x}_{i}(t)\big)^{b}\Big]}{\Big[w+\mathbf{x}_{i}^{b}(t)\mathbf{s}_{i}(t)+\big(1-\mathbf{x}_{i}(t)\big)^{b}\big(d_{i}-\mathbf{s}_{i}(t)\big)\Big]^{2}}
\nonumber\\
= &
\frac{\Big[\mathbf{x}_{i}^{b}(t)\big(1-\mathbf{x}_{i}(t)\big)+\mathbf{x}_{i}(t)\big(1-\mathbf{x}_{i}(t)\big)^{b}\Big]w+\mathbf{x}_{i}^{b}(t)\big(1-\mathbf{x}_{i}(t)\big)^{b}d_{i}}{\Big[w+\mathbf{x}_{i}^{b}(t)\mathbf{s}_{i}(t)+\big(1-\mathbf{x}_{i}(t)\big)^{b}\big(d_{i}-\mathbf{s}_{i}(t)\big)\Big]^{2}}
\nonumber\\
> & 0,
\end{align}
when $\mathbf{x}_{i}(t)\in (0,1)$.
 Due to (i) and (\ref{partial}), we obtain when $\mathbf{s}_{i}(t)>\mathbf{s}^{*}\big(\mathbf{x}_{i}(t),b\big)d_{i}$, $\mathbf{x}_{i}(t+1)>\mathbf{x}_{i}(t)$ holds.

\noindent(iii). the statement follows from the same analysis as in the proof of (ii).

The desired lemma thus holds. \hfill$\Box$

\subsection{Proofs of Statements}

\subsubsection{Proof of Proposition \ref{1/2 unstable}}
When $\mathbf{x}_{i}(t)=1/2$ for all $i=1,2,...,n$, we know that $\mathbf{s}_{i}(t)=d_{i}/2$ for all $i\in \mathrm{V}$. Thus,
\begin{align*}
\mathbf{x}_{i}(t+1)=\frac{w/2+(1/2)^{b}d_{i}/2}{w+(1/2)^{b}d_{i}/2+(1/2)^{b}d_{i}/2}=1/2
\end{align*}
for all $i\in \mathrm{V}$.
Therefore, we have proved that $\mathsf{X}=(1/2,1/2,...,1/2)^\top$ is an equilibrium.

Next, we show that $\mathsf{X}=(1/2,1/2,...,1/2)^\top$ is unstable.
Let
\begin{align*}
\mathbf{X}(0)=(1/2-\theta,1/2-\theta,...,1/2-\theta)^\top
\end{align*}
where $\theta \in (0,1/2)$. From Theorem \ref{theorem1}, there holds $\lim_{t\to \infty}\mathbf{X}(t)=(0,0,...,0)^\top$. It is clear from this point $(1/2,1/2,...,1/2)^\top$ cannot be a stable equilibrium.
We have proved the desired result. \hfill$\Box$

\subsubsection{Proof of Theorem \ref{complete equilibria}}

From the definitions of complete graph and $\mathbf{s}^{*}(x,b)$, there hold
\begin{align}\label{n-1}
\frac{\sum_{k=1,k\neq i}^{n}\mathsf{x}_{k}}{n-1}=\mathbf{s}^{*}(\mathsf{x}_{i},b) =\frac{(1-\mathsf{x}_{i})^{b-1}}{\mathsf{x}_{i}^{b-1}+(1-\mathsf{x}_{i})^{b-1}}
\end{align}
and
\begin{align*}
\sum_{k=1}^{n}\mathsf{x}_{k}=\sum_{k=1,k\neq i}^{n}\mathsf{x}_{k}+\mathsf{x}_{i}=\frac{(n-1)(1-\mathsf{x}_{i})^{b-1}}{\mathsf{x}_{i}^{b-1}+(1-\mathsf{x}_{i})^{b-1}}+
\mathsf{x}_{i}
\end{align*}
for all $i\in \mathrm{V}$.
Let
\begin{align*}
g_{b}(x)=\frac{(n-1)(1-x)^{b-1}}{x^{b-1}+(1-x)^{b-1}}+x
\end{align*}
for $x\in(0,1)$. This immediately gives us that  $\sum_{k=1}^{n}\mathsf{x}_{k}=g_{b}(\mathsf{x}_{i})$
for all $i\in \mathrm{V}$, and
\begin{align*}
\frac{d}{dx}g_{b}(x)=1-\frac{(n-1)(b-1)(1-x)^{b-2}x^{b-2}}{\big[x^{b-1}+(1-x)^{b-1}\big]^{2}}.
\end{align*}

When $x\in (0,1)$ and $b\leq 1$, we conclude  that
\begin{align*}
\frac{d}{dx}g_{b}(x)>0
\end{align*}
 and thus $g_{b}(x)$
is monotonic. Consequently, in view of $g_{b}(\mathsf{x}_{i})=\sum_{k=1}^{n}\mathsf{x}_{k}=g_{b}(\mathsf{x}_{j})$, there holds that  $\mathsf{x}_{i}=\mathsf{x}_{j}$ for all $i,j \in \mathrm{V}$ when $\mathsf{X}\in \mathbb{E}_{\rm int}$.
Hence,  we can assume $\mathsf{x}_{i}=\tilde{x}$ for all $i\in \mathrm{V}$.
According to (\ref{n-1}) we know
\begin{align*}
\tilde{x}= \frac{(1-\tilde{x})^{b-1}}{\tilde{x}^{b-1}+(1-\tilde{x})^{b-1}},
\end{align*}
which implies $\tilde{x}=1/2$.
We have now obtained that if $b\leq 1$, the only interior equilibrium is  $(1/2,1/2,...,1/2)^\top$.
When $x\in (0,1)$ and $b=2$, we have
\begin{align*}
\frac{d}{dx}g_{b}(x)<0.
\end{align*}
The fact that $(1/2,1/2,...,1/2)^\top$ is the unique equilibrium can be established using a similar analysis.
Finally, the instability can be deduced from Proposition \ref{1/2 unstable} directly. Now we have completed the proof.
\hfill$\Box$

\subsubsection{Proof of Theorem \ref{star equilibria}}

(i). According to the definitions of $\mathbf{s}^{*}(x,b)$ and star graph, we obtain
\begin{align*}
\begin{split}
\mathbf{s}^{*}(\mathsf{x}_{i},b) ={} & \mathsf{x}_{n}, i=1,2,...,n-1; \\
\mathbf{s}^{*}(\mathsf{x}_{n},b) ={} & \frac{\sum_{i=1}^{n-1}\mathsf{x}_{i}}{n-1}.
\end{split}
\end{align*}
When $b=1$, for all $i \in \mathrm{V}\backslash \{n\}$, there holds
\begin{align}\label{b=1}
\mathsf{x}_{i}=\frac{w\mathsf{x}_{i}+\mathsf{x}_{i}\mathsf{x}_{n}}{w +\mathsf{x}_{i}\mathsf{x}_{n}+(1-\mathsf{x}_{i})(1-\mathsf{x}_{n})}.
\end{align}
In view of $\mathsf{x}_{i} \neq 0$ or $1$ for $i=1,2,...,n-1$, (\ref{b=1}) immediately gives us $\mathsf{x}_{n}= 1/2$. Besides, we have
\begin{align}\label{star1}
\mathbf{s}^{*}(1/2,1)=\frac{1}{2}=\frac{\sum_{i=1}^{n-1}\mathsf{x}_{i}}{n-1}.
\end{align}
From (\ref{star1}) it is easy to verity  when $b=1$, $\mathbb{E}_{\rm int}=\big\{(\mathsf{x}_{1},\mathsf{x}_{2},...,\mathsf{x}_{n-1},1/2)^\top: \sum_{i=1}^{n-1}
\mathsf{x}_{i}=(n-1)/2,\mathsf{x}_{i}\in (0,1) \text{ for all } i\in \mathrm{V} \backslash \{n\} \big\}$.

Next, we prove the instability of any equilibrium $\mathsf{X}\in \mathbb{E}_{\rm int}$.
For any equilibrium $\mathsf{X}=(\mathsf{x}_{1},\mathsf{x}_{2},...,\mathsf{x}_{n-1},1/2)$ where $\sum_{i=1}^{n-1}\mathsf{x}_{i}=(n-1)/2$ and $\mathsf{x}_{i}\in (0,1)$ for all $i\in \mathrm{V} \setminus \{n\}$, let
$\mathbf{x}_{i}(0)\in[0,\mathsf{x}_{i})$ for all $i\in\mathrm{V}$. Then we will prove that $\mathbf{x}_{i}(t)$ is decreasing when for all $i\in\mathrm{V}$ there holds that $\mathbf{x}_{i}(t)\in[0,\mathsf{x}_{i})$.
We see
\begin{align}\label{xnt}
\mathbf{x}_{n}(t+1)-\mathbf{x}_{n}(t)
=
\frac{\mathbf{x}_{n}(t)\big(1-\mathbf{x}_{n}(t)\big)\big(2\mathbf{s}_{n}(t)-d_{n}\big)}{w+\mathbf{x}_{n}(t)\mathbf{s}_{n}(t)+\big(1-\mathbf{x}_{n}(t)\big)\big(d_{n}-\mathbf{s}_{n}(t)\big)}
\leq
0,
\end{align}
where the inequation holds because $2\mathbf{s}_{n}(t)=2\sum_{i=1}^{n-1}\mathbf{x}_{i}(t)<2\times\frac{n-1}{2}=d_{n}$.
Similiarly, for any $i\in \{1,2,...,n-1\}$, we obtain
\begin{align}\label{xit}
\mathbf{x}_{i}(t+1)-\mathbf{x}_{i}(t)
=
\frac{\mathbf{x}_{i}(t)\big(1-\mathbf{x}_{i}(t)\big)\big(2\mathbf{x}_{n}(t)-d_{i}\big)}{w+\mathbf{x}_{i}(t)\mathbf{x}_{n}(t)+\big(1-\mathbf{x}_{i}(t)\big)\big(d_{n}-\mathbf{x}_{n}(t)\big)}
\leq
0.
\end{align}
Notice that $\mathbf{x}_{i}(0)\in[0,\mathsf{x}_{i}),i\in\mathrm{V}$, we thus know $\mathbf{x}_{i}(t)\in[0,\mathsf{x}_{i})$ is decreasing for all $i=1,2,...,n$ and $t>0$.

Let $\mathbf{z}(t)=\min_{i\in\mathrm{V}}\big\{\mathbf{x}_{i}(t)\big\}$. Then we have that $\mathbf{z}(t)$ is decreasing. From $\mathbf{z}(t)\ge 0$, there holds $\lim_{t\to \infty}\mathbf{z}(t)=\mathbf{z}^{*}\ge 0$. Next we will prove that $\mathbf{z}^{*}=0$.
We can conclude the following results easily
\begin{align}\label{bound}
\mathbf{z}^{*} \leq \mathbf{x}_{i}(t)\leq \mathbf{x}_{i}(0)<\mathsf{x}_{i}
\end{align}
for all $i\in\mathrm{V}$.
Due to (\ref{xnt}) and (\ref{bound}), we obtain
\begin{align*}
\big\lvert \mathbf{x}_{n}(t+1)-\mathbf{x}_{n}(t) \big\lvert \ge \frac{\mathbf{z}^{*}\big(1-\mathbf{x}_{n}(0)\big)\big((n-1)-\sum_{i}^{n-1}\mathbf{x}_{i}(0)\big)}{w+\frac{1}{2}\frac{n-1}{2}+(1-\mathbf{z}^{*})^{2}(n-1)}=c_{n}
\end{align*}
where $c_{n}$ is a constant determined by $w, \mathbf{z}^{*}$ and $\mathbf{x}_{i}(0),i\in\mathrm{V}$. When $\mathbf{z}^{*}>0$, we have that $c_{n}> 0$  and then $\lim_{t\to \infty}\mathbf{x}_{n}(t)=-\infty<0$ monotonously. This immediately gives us a contradiction. Therefore, we know that $\mathbf{z}^{*}=0$. It means that there exists $j\in \mathrm{V}$ such that $\lim_{t\to \infty}\mathbf{x}_{j}(t)=0$.
From the monotonous decrease of $\mathbf{x}_{j}(t)$, we know that when $\varepsilon=\frac{1}{2}\min_{i\in \mathrm{V}}\mathsf{x}_{i}>0$, there exists $N\in\mathbb{N}$ such that $\mathbf{x}_{j}(t)<\varepsilon$ and $\big\lVert \mathbf{X}(t)-\mathsf{X}\big\Vert\ge\big|\mathbf{x}_{j}(t)-\mathsf{x}_{j}\big|>\frac{1}{2}\min_{i\in \mathrm{V}}\mathsf{x}_{i}=\varepsilon$ when $t>N$.
For any small enough $\delta>0$, let  $\mathbf{x}_{i}(0)=\mathsf{x}_{i}-\frac{\delta}{n}\in[0,\mathsf{x}_{i})$ and then $\big\lVert \mathbf{X}(0)-\mathsf{X} \big\Vert=\frac{\delta}{\sqrt{n}}<\delta$. But $\lim_{t\to\infty}\big\lVert \mathbf{X}(t)-\mathsf{X}\big\Vert>\varepsilon$.
This immediately gives us that when $b=1$ such equilibria are unstable from the definition of unstability.

\noindent(ii). When $b\neq 1$, for all $i \in \mathrm{V}\backslash \{n\}$, there holds
\begin{align*}
\mathsf{x}_{i}=\frac{w \mathsf{x}_{i}+\mathsf{x}_{i}^{b}\mathsf{x}_{n}}{w +\mathsf{x}_{i}^{b}\mathsf{x}_{n}+(1-\mathsf{x}_{i})^{b}(1-\mathsf{x}_{n})}.
\end{align*}
Because $b\neq 1$ and $\mathsf{x}_{i}\in (0,1)$ for all $i\in\mathrm{V}$, we conclude that
\begin{align*}
w\mathsf{x}_{i} +\mathsf{x}_{i}^{b+1}\mathsf{x}_{n}+\mathsf{x}_{i}(1-\mathsf{x}_{i})^{b}(1-\mathsf{x}_{n}) = w \mathsf{x}_{i}+\mathsf{x}_{i}^{b}\mathsf{x}_{n},
\end{align*}
which implies
\begin{align*}
\mathsf{x}_{i}= \frac{1}{(\frac{\mathsf{x}_{n}}{1-\mathsf{x}_{n}})^{1/(b-1)}+1}
\end{align*}
for all $i=1,2,...,n-1$. Therefore, $\mathsf{x}_{i}=\mathsf{x}_{1}$ for all $i\in \mathrm{V} \backslash \{n\}$.

We now know that
\begin{align*}
\begin{split}
\mathsf{x}_{1}=\frac{(n-1) \mathsf{x}_{1}}{(n-1)}=\mathbf{s}^{*}(\mathsf{x}_{n},b)
= {} &
\frac{(1-\mathsf{x}_{n})^{b-1}}{(1-\mathsf{x}_{n})^{b-1}+\mathsf{x}_{n}^{b-1}},
\\
 \mathsf{x}_{n}=\mathbf{s}^{*}(\mathsf{x}_{1},b)
 = {} &
 \frac{(1-\mathsf{x}_{1})^{b-1}}{(1-\mathsf{x}_{1})^{b-1}+\mathsf{x}_{1}^{b-1}}
.
\end{split}
\end{align*}
Therefore, we obtain
\begin{align}\label{x1+xn=1}
\begin{split}
\frac{\mathsf{x}_{1}}{1-\mathsf{x}_{1}}
= {} &
\Big(\frac{1-\mathsf{x}_{n}}{\mathsf{x}_{n}}\Big)^{b-1},
\\
 \frac{\mathsf{x}_{n}}{1-\mathsf{x}_{n}}
= {} &
\Big(\frac{1-\mathsf{x}_{1}}{\mathsf{x}_{1}}\Big)^{b-1}
.
\end{split}
\end{align}
If $b=2$, it must be the case that $\mathsf{x}_{1}+\mathsf{x}_{n}=1$ where $\mathsf{x}_{1},\mathsf{x}_{n}\in (0,1)$ from (\ref{x1+xn=1}). Now, we can verity  for $b=2$ $\mathbb{E}_{\rm int}=\big\{(\mathsf{x}_{1},\mathsf{x}_{1},...,\mathsf{x}_{1},1-\mathsf{x}_{1})^\top: \mathsf{x}_{1} \in (0,1) \big\}$.

Now, we prove the instability of any equilibrium $\mathsf{X}\in \mathbb{E}_{\rm int}$ when $b=2$.
For the equilibrium $\mathsf{X}=(a,a,...,a,1-a)^\top$ where $a\in (0,1)$, let $\mathbf{x}_{1}(0)=\mathbf{x}_{2}(0)=...=\mathbf{x}_{n-1}(0)\in[0,a)$ and $\mathbf{x}_{n}(0)\in[0,1-a)$.
We show that $\mathbf{x}_{1}(t)$ is  decreasing when for all $i\in\mathrm{V}$ there holds that $\mathbf{x}_{i}(t)\in[0,\mathsf{x}_{i})$. We have
\begin{align*}
\mathbf{x}_{1}(t+1)-\mathbf{x}_{1}(t)
=
\frac{\mathbf{x}_{1}(t)\big(1-\mathbf{x}_{1}(t)\big)\Big[\mathbf{x}_{1}(t)\mathbf{x}_{n}(t)-\big(1-\mathbf{x}_{1}(t)\big)\big(1-\mathbf{x}_{n}(t)\big)\Big]}{w+\mathbf{x}_{1}^{2}(t)\mathbf{x}_{n}(t)+\big(1-\mathbf{x}_{1}(t)\big)^{2}\big(1-\mathbf{x}_{n}(t)\big)}
\leq
0,
\end{align*}
where the inequation holds because $0<\mathbf{x}_{1}(t)<a<1-\mathbf{x}_{n}(t)$ and $0<\mathbf{x}_{n}(t)<1-a<1-\mathbf{x}_{1}(t)$.
Analogously, we have that  for all $i\in\mathrm{V}$, $\mathbf{x}_{i}(t)$ is decreasing when $\mathbf{x}_{1}(0)=...=\mathbf{x}_{n-1}(0)\in[0,a)$ and $\mathbf{x}_{n}(0)\in[0,1-a)$. Then we can also
prove that
$\lim_{t\to\infty}\min_{i\in\mathrm{V}}\big\{\mathbf{x}_{i}(t)\big\}=0$ and
 $\lim_{t\to\infty}\big\lVert \mathbf{X}(t)-\mathsf{X}\big\Vert>\frac{1}{2}\min\{a,1-a\}$ for any $\mathbf{X}(0)$ satisfying that $\mathbf{x}_{1}(0)=...=\mathbf{x}_{n-1}(0)\in[0,a)$ and $\mathbf{x}_{n}(0)\in[0,1-a)$.
We thus obtain that $\mathsf{X}$ is unstable similarly to the case in (i).

\noindent(iii). From (\ref{x1+xn=1}), there holds
 \begin{align}\label{2-node equation}
\frac{\mathsf{x}_{1}}{1-\mathsf{x}_{1}}
=
\Big(\frac{\mathsf{x}_{1}}{1-\mathsf{x}_{1}}\Big)^{(b-1)^{2}}
\end{align}
when $b\neq 1$. Recall that $b>0$ and $b \neq 2$, we obtain $(b-1)^{2}\neq 1$ and
the solutions of equation (\ref{2-node equation}) are given by  $\frac{\mathbf{x}_{1}}{1-\mathbf{x}_{1}}$ be either $0$ or $1$.
Due to $\mathsf{x}_{1}=...=\mathsf{x}_{n-1}\in(0,1)$, this immediately gives us that $\mathsf{x}_{1}=...=\mathsf{x}_{n-1}=1/2$. We can therefore readily conclude that $\mathsf{x}_{n}=1/2$ from (\ref{x1+xn=1}). Thus when $b\neq1$ or $2$, the only interior equibrium is $(1/2,1/2,...,1/2)^\top$.
From Proposition \ref{1/2 unstable}, there holds that $(1/2,...,1/2)^\top$ is unstable.

We therefore have completed the proof. \hfill$\Box$

\subsubsection{Proof of Theorem \ref{cycle equilibria}}
(i). In view of Lemma \ref{lemma 1} and the definition of cycle graph, we have
\begin{align}\label{cycle}
\frac{\mathsf{x}_{i-1}+\mathsf{x}_{i+1}}{2}=\mathbf{s}^{*}(\mathsf{x}_{i},b)=\frac{(1-\mathsf{x}_{i})^{b-1}}{(1-\mathsf{x}_{i})^{b-1}+\mathsf{x}_{i}^{b-1}},
\end{align}
for all $i\in\mathrm{V}$.
When $b=1$ and $n\equiv 1 (mod \ 4)$, according to (\ref{cycle}), we know
\begin{align}\label{cycle b=1}
\mathsf{x}_{i}+\mathsf{x}_{i+2}=1 \text{ and }
 \mathsf{x}_{i+2}+\mathsf{x}_{i+4}=1,
\end{align}
for all $i\in\mathrm{V}$. Therefore, we obtain that $\mathsf{x}_{i}=\mathsf{x}_{i+4k}$ for all $i\in\mathrm{V}$ and $k\in \mathbb{Z}$.
Due to $n\equiv 1 (mod \ 4)$, let $n=4m+1$ where $m\in \mathbb{N}$. Then we have that $\mathsf{x}_{i}=\mathsf{x}_{i+4m}=\mathsf{x}_{i-1}$ for all $i\in\mathrm{V}$. This gives us that $\mathsf{x}_{1}=\mathsf{x}_{2}=...=\mathsf{x}_{n}$. From (\ref{cycle b=1}), $\mathsf{x}_{i}=1/2$ holds for all $i\in\mathrm{V}$. The only equilibrium is $(1/2,1/2,...,1/2)^\top$ when $b=1$ and $n\equiv 1 (mod \ 4)$.
If $n=4m+2$ or $n=4m+3$, the results can be obtained analogously. From Proposition \ref{1/2 unstable}, there holds that $(1/2,...,1/2)^\top$ is unstable.

\noindent(ii). When $b=1$ and  $n\equiv 0 (mod \ 4)$, let
\begin{align*}
\begin{split}
\mathsf{x}_{i}= {} & a_{1}, \text{when } i\equiv 0(mod \ 4);
\\
\mathsf{x}_{i}= {} & a_{2}, \text{when } i\equiv 1(mod \ 4);
\\
\mathsf{x}_{i}= {} & a_{3}, \text{when } i\equiv 2(mod \ 4);
\\
\mathsf{x}_{i}= {} & a_{4}, \text{when } i\equiv 3(mod \ 4)
\end{split}
\end{align*}
where $a_{1},a_{2},a_{3},a_{4}\in (0,1)$ for all $i\in\mathrm{V}$. Noting  (\ref{cycle b=1}), there hold  $a_{1}+a_{3}=1$ and $a_{2}+a_{4}=1$. Now, we can verity  when $b=1$ and  $n\equiv 0 (mod \ 4)$ all interior equilibria are $\mathsf{X}=(a_{1},a_{2},1-a_{1},1-a_{2},a_{1},...,1-a_{1},1-a_{2})^\top$ where $a_{1},a_{2} \in (0,1)$.

Next, we prove that when $b=1$ and $n\equiv 0 (mod \ 4)$, the equilibrium $\mathsf{X}=(a_{1},a_{2},1-a_{1},1-a_{2},...,1-a_{2})^\top$ where $a_{1},a_{2} \in (0,1)$ is unstable.
Suppose $\mathbf{x}_{i+4k}(0)=\mathbf{x}_{i}(0)$ for all $k\in\mathbb{N}$ and $i=1,2,3,4$. Besides, let $\mathbf{x}_{1}(0)<a_{1},\mathbf{x}_{2}(0)<a_{2},\mathbf{x}_{3}(0)<1-a_{1},\mathbf{x}_{4}(0)<1-a_{2}$.
We will prove that $\mathbf{x}_{i}(t)$ is decreasing for all $i=1,2,3$ and $4$.
We first show  that $\mathbf{x}_{1}(t)$ is decreasing when $\mathbf{x}_{i}(t)\in[0,\mathsf{x}_{i})$ for all $i\in\mathrm{V}$. There holds
\begin{align*}
\mathbf{x}_{1}(t+1)-\mathbf{x}_{1}(t)
=
\frac{2\mathbf{x}_{1}(t)\big(1-\mathbf{x}_{1}(t)\big)\big(\mathbf{s}_{1}(t)-1\big)}{w+\mathbf{x}_{1}(t)\mathbf{s}_{1}(t)+\big(1-\mathbf{x}_{1}(t)\big)\big(2-\mathbf{s}_{1}(t)\big)}
\leq
0,
\end{align*}
where the inequation holds because $\mathbf{s}_{1}(t)=\mathbf{x}_{2}(t)+\mathbf{x}_{n}(t)<a_{2}+1-a_{2}<1$. Analogously,  $\mathbf{x}_{i}(t),i=2,3,4$ are decreasing  when $\mathbf{x}_{i}(t)\in[0,\mathsf{x}_{i})$ for all $i\in\mathrm{V}$. Therefore, we obtain that $\mathbf{x}_{i}(t)$  is decreasing when $\mathbf{x}_{i}(0)=\mathbf{x}_{i+4k}(0)\in[0,\mathsf{x}_{i}), i=1,2,3,4,k\in\mathbb{N}$.
Then we can also prove that
$\lim_{t\to\infty}\min_{i\in\mathrm{V}}\big\{\mathbf{x}_{i}(t)\big\}=0$ and $\lim_{t\to\infty}\big\lVert \mathbf{X}(t)-\mathsf{X}\big\Vert>\frac{1}{2}\min\{a_{1},a_{2},1-a_{1},1-a_{2}\}$ for any $\mathbf{X}(0)$ satisfying that $\mathbf{x}_{i}(0)=\mathbf{x}_{i+4k}(0)\in[0,\mathsf{x}_{i}), i=1,2,3,4,k\in\mathbb{N}$.
We thus obtain that $\mathsf{X}=(a_{1},a_{2},1-a_{1},1-a_{2},...,1-a_{2})^\top$ where $a_{1},a_{2} \in (0,1)$ is unstable similarly to the proof of Theorem \ref{star equilibria}.
Therefore, we prove that such equilibria are unstable.

\noindent(iii). We now discuss the interior equilibria when $b=2$. Let $\mathsf{y}_{i}=\mathsf{x}_{i}+\mathsf{x}_{i+1}$ for all $i\in\mathrm{V}$.
From (\ref{cycle}) and $b=2$, we obtain
\begin{align*}
\frac{\mathsf{x}_{i-1}+\mathsf{x}_{i+1}}{2}=1-\mathsf{x}_{i},
\end{align*}
for all $i\in\mathrm{V}$. Therefore, there holds
\begin{align*}
\mathsf{y}_{i}+\mathsf{y}_{i+1}=2,
\end{align*}
for all $i\in\mathrm{V}$. Accordingly $\mathsf{y}_{i}=\mathsf{y}_{i+2k}$ holds for all $i\in \mathrm{V}$ and $k\in \mathbb{Z}$.
When $n=2m+1$ where $m\in \mathbb{N}$, we obtain that $\mathsf{y}_{i}=\mathsf{y}_{i+2m}=\mathsf{y}_{i-1}$ for all $i\in \mathrm{V}$. Therefore, we have that $\mathsf{y}_{i}=1$ for all  $i\in \mathrm{V}$. That is,
\begin{align*}
\mathsf{x}_{i}+\mathsf{x}_{i+1}=1,
\end{align*}
for all $i\in\mathrm{V}$. Because $n=2m+1$ where $m\in \mathbb{N}$, we see that $\mathsf{x}_{i}=\mathsf{x}_{i+2m}=\mathsf{x}_{i-1}$ for all $i\in \mathrm{V}$. This immediately gives us $\mathsf{x}_{1}=\mathsf{x}_{2}=...=\mathsf{x}_{n}=1/2$. From Proposition \ref{1/2 unstable}, there holds that $(1/2,...,1/2)^\top$ is unstable.

\noindent(iv). When $b=2$ and $n=2m$ where $m\in \mathbb{N}$, let
\begin{align*}
\begin{split}
\mathsf{y}_{i}= {} & \tilde{a}, \text{when } i\equiv 0(mod \ 2);
\\
\mathsf{y}_{i}= {} & 2-\tilde{a}, \text{when } i\equiv 1(mod \ 2)
\end{split}
\end{align*}
where $\tilde{a}\in (0,2)$ for all $i\in\mathrm{V}$.
In view of the definition of $\mathsf{y}_{i}$, we know that $\sum_{i=1}^{n}\mathsf{x}_{i}=\sum_{i=1}^{m}\mathsf{y}_{2i-1}=\sum_{i=1}^{m}\mathsf{y}_{2i}$. Therefore, we get that $m\tilde{a}=m(2-\tilde{a})$ and $\tilde{a}=1$. This tolds us
\begin{align*}
\mathsf{x}_{i}+\mathsf{x}_{i+1}=1
\end{align*}
for all $i\in\mathrm{V}$. Because $n=2m$, we know that $\mathsf{x}_{i}=\mathsf{x}_{i+2k}$ for all $i\in\mathrm{V}$ and $k\in \mathbb{N}$. We thus conclude that $\mathbb{E}_{\rm int}=\big\{(a,1-a,a,1-a,...,a,1-a)^\top:a \in (0,1) \big\}$ when $b=2$ and $n\equiv 0 (mod \ 2)$.
The instability of $(a,1-a,a,1-a,...,a,1-a)^\top$ can be proved similarly to the statement (ii) in Theorem \ref{star equilibria}.
This completes the proof. \hfill$\Box$

\section{Conclusions}\label{sec:conclusions}

We have provided  a systemic analysis to   social opinion dynamics subject to individual biases, which generated state-dependent edge weights and therefore   highly nonlinear network dynamics. It was shown that when the initial network opinions are polarized towards one side of the state space,  node biases would drive the opinion evolution to the corresponding interval boundaries under quite general network conditions. For a few fundamental network structures, some important  interior network equilibria were presented   for a wide range of system parameter in terms of their positions and stabilities, where the interval centroid was proven to be unstable regardless of the bias level and the network topologies. Future work includes studies of the distribution  and stability of equilibria under more general network structures, especially those that are resilient subject to network structure switches as such structure change is common for real-world social networks.

\end{document}